\pdfoutput=1
\documentclass[superscriptaddress, 12pt, aps, prl,  preprint, longbibliography]{revtex4-1}
\usepackage{graphicx}
\usepackage{epstopdf}
\usepackage{color}
\usepackage{amsmath, amssymb}
\usepackage{bm}
\usepackage{float}
\usepackage{placeins}
\usepackage{physics}
\usepackage{multirow}
\begin{document}

\title{Raman spectra of hydrocarbons under extreme conditions of pressure and temperature: a first-principles study}
\author{Rui Hou}
\affiliation{Department of Physics, Hong Kong University of Science and Technology, Hong Kong, China}
\author{Ding Pan}
\email{dingpan@ust.hk}
\affiliation{Department of Physics and Department of Chemistry, Hong Kong University of Science and Technology, Hong Kong, China}
\affiliation{HKUST Fok Ying Tung Research Institute, Guangzhou, China}

\date{\today}

\begin{abstract}
Hydrocarbons are of great importance in carbon-bearing fluids in deep Earth and in ice giant planets at extreme pressure (P)-temperature (T) conditions. Raman spectroscopy is a powerful tool to study the chemical speciation of hydrocarbons; however, it is challenging to interpret Raman data at extreme conditions.
Here, we performed ab initio molecular dynamics simulations coupled with the modern theory of polarization to calculate Raman spectra of methane, ethane, and propane up to 48 GPa and 2000 K.
Our method includes anharmonic temperature effects.
We studied the pressure and temperature effects on the Raman bands, and identified the characteristic Raman modes for the C-C and C-C-C bonds. 
Our result may help to interpret in-situ Raman data of hydrocarbons at extreme P-T conditions, with important implications for understanding the deep carbon cycle inside Earth and the compositions of ice giant planets.

\end{abstract}

\maketitle

\section{introduction}
Methane (CH$_4$) and other light hydrocarbons are important components of reduced carbon-hydrogen-oxygen (C-H-O) fluids inside deep Earth \cite{manning2013chemistry, sephton2013origins}. Methane is also a major constituent of deep interiors of two ice giant planets in the solar system:  Uranus and Neptune \cite{Hubbard145, podolak_comparative_1995}. 
The abiotic hydrocarbons may participate in the formation of petroleum deposits; this controversial hypothesis can be dated back to the time of Dimitri Mendeleev \cite{Mendeleev1877,sephton2013origins}.
The properties of hydrocarbons under extreme pressure (P) and temperature (T) conditions play crucial roles in the deep carbon cycle in Earth \cite{sephton2013origins}, and in shaping the structure and dynamics of ice giant planets.
Previously, it was assumed that hydrocarbons were simply mixed with other small volatile molecules, e.g., H$_2$O, CO$_2$, H$_2$, in C-H-O fluids \cite{zhang2009model}, or with H$_2$O and NH$_3$ inside Uranus and Neptune \cite{podolak_comparative_1995}. 
However, many recent theoretical and experimental studies show that methane and other hydrocarbons may further react at extreme P-T conditions
(e.g., \cite{ ancilotto_dissociation_1997,hirai_polymerization_2009,kolesnikov_methane-derived_2009,gao_dissociation_2010,Spanu2011,lobanov_carbon_2013}). 
Using ab initio molecular dynamics (AIMD) simulations,
Ancilotto et al. found that at 100 GPa and 4000 K, methane dissociates into a mixture of methane, ethane (C$_2$H$_6$), and propane (C$_3$H$_8$) \cite{ancilotto_dissociation_1997}.
Spanu et al. predicted that higher hydrocarbons become thermodynamically more favored than methane at above 4 GPa and 1000$\sim$2000 K \cite{spanu_stability_2011}. 
Experiments with laser-heated diamond anvil cells reported that ethane, propane, and higher hydrocarbons may form out of methane at pressures higher than 2 GPa and temperatures above 1000 K
\cite{kolesnikov_methane-derived_2009, hirai_polymerization_2009}. 

Raman spectroscopy is the most widely used experimental tool to detect chemical speciation in C-H-O fluids at \emph{both} high-P \emph{and} high-T conditions \cite{goncharov2012raman}.
Raman spectra of vibrations provide fingerprints to identify molecules and ions in in-situ experiments \cite{goncharov2012raman, doi:10.1002/9781119961284.ch1}.
However, it is very challenging to measure Raman spectra under extreme P-T conditions, because many high
energy levels are excited at high temperatures and the signals from diamond anvils may
interfere; for example,
the peaks near 1332 cm$^{-1}$ and 2400 cm$^{-1}$ are
attributed to the first- and the second-order scattering from diamond anvils, respectively \cite{goncharov2005dynamic}.
Thus, theoretically studying the P-T dependence of Raman bands is of great importance to the interpretation of Raman spectra under extreme conditions \cite{lobanov_carbon_2013, petrov_raman_2018}.
In first-principles calculations, we usually calculate the change of electronic polarizability along a normal vibrational mode under the harmonic approximation to obtain the Raman spectra of hydrocarbons in the gas or solid phase (e.g., \cite{lin2007correlation, lin_structural_2011, ramya2012raman}), where the temperature is treated as 0 K; however, for C-H fluids at the supercritical state, 
the anharmonic and temperature effects can not be ignored \cite{putrino2002anharmonic, wan_raman_2013,pan2020first}.

Here, we performed AIMD simulations coupled with density functional perturbation theory to calculate the Raman spectra of methane, ethane, and propane at 13.4$\sim$48 GPa and 1445$\sim$2000 K.
The electronic polarizability was calculated on the fly along with AIMD simulations. 
We studied how pressure and temperature affect Raman spectra, and assigned Raman bands at the molecular scale. We discussed possible characteristic Raman signals for the formation of higher hydrocarbons.
Our work has important implications for studying chemical speciation of hydrocarbons at extreme P-T conditions found in deep Earth and ice giant planets.

\section{Methods}
We carried out AIMD simulations using the Qbox package \cite{gygi_architecture_2008, qbox_url} with the Born–Oppenheimer approximation and a time step of 5 a.u. 
We used the Perdew–Burke–Ernzerhof (PBE) exchange-correlation functional \cite{perdew_pbe} and the HSCV norm-conserving pseudopotentials \cite{hamann_norm-conserving_1979, vanderbilt_optimally_1985, pps}. The kinetic energy cutoff of plane waves is 65 Ry, which was increased to 145 Ry to calculate pressure. 
We applied the Bussi-Donadio-Parrinello thermostat to control temperature \cite{bussi_canonical_2007}.
For methane, ethane, and propane, 
the simulation boxes with periodic boundary conditions 
contain 56, 32, and 23 molecules, respectively.

We performed on-the-fly calculations using density functional perturbation theory (DFPT) \cite{baroni_phonons_2001} implemented in the Qbox package along with AIMD simulations to calculate electronic polarizability, which were conducted every 25 MD steps after about 10 ps equilibration. Each MD trajectory is longer than 130 ps. We computed Raman spectra of supercritical fluids using the Fourier transfer of the autocorrelation functions of polarizabilities.
The isotropic and anisotropic Raman spectra are calculated respectively by
\begin{align}
    R_{\mathrm{iso}}(\omega) \propto \frac{\hbar \omega}{k_{B} T} \int d t e^{-i \omega t}\frac{\left\langle \bar{\alpha}^{\mathrm{}}(0) \bar{\alpha}^{\mathrm{}}(t)\right\rangle}{\left\langle \bar{\alpha}^{\mathrm{}}(0) \bar{\alpha}^{\mathrm{}}(0)\right\rangle}
    \label{eq:isotropic}\\
    R_{\mathrm{aniso}}(\omega) \propto \frac{2}{15} \frac{\hbar \omega}{k_{B} T} \int d t e^{-i \omega t}\frac{\left\langle \Tr  (\beta^{\mathrm{}}(0) \beta^{\mathrm{}}(t))\right\rangle}{\left\langle\Tr  (\beta^{\mathrm{}}(0) \beta^{\mathrm{}}(0))\right\rangle},
    \label{eq:aninstropic}
\end{align}
where $\hbar$ is the Planck constant, $\omega$ is the Raman frequency, $k_B$ is the Boltzmann constant, Tr stands for the trace of matrix. $\bar{\alpha}$ and $\beta$ denote the isotropic and anisotropic components of the polarizability tensor $\alpha$, $\bar{\alpha}=\frac{1}{3}Tr\alpha$ and $\beta=\alpha-\bar{\alpha}I$, respectively.
In high-P and high-T experiments, we measured unpolarized Raman spectra, which are obtained from the linear combination of Eqs. (\ref{eq:isotropic}) and (\ref{eq:aninstropic}): 
\begin{equation}
    R_{unpol}= R_{iso} + \frac{7}{4}R_{aniso}.
\end{equation}
We further smoothed Raman spectra using Gaussian broadening, whose full width at half maximum is 25 cm$^{-1}$. 

For the gas and solid phases, DFPT calculations were carried out using the Quantum Espresso package\cite{Giannozzi_2009}.
For methane, ethane, and propane in the gas phase, 
the mean absolute error of calculated Raman frequencies is 23 cm$^{-1}$ compared with experimental values, indicating that our computational setups are reliable.

\section{Results and Discussion}

\begin{figure}
    \centering
    \vspace{5mm}
    \includegraphics[width=0.6\textwidth]{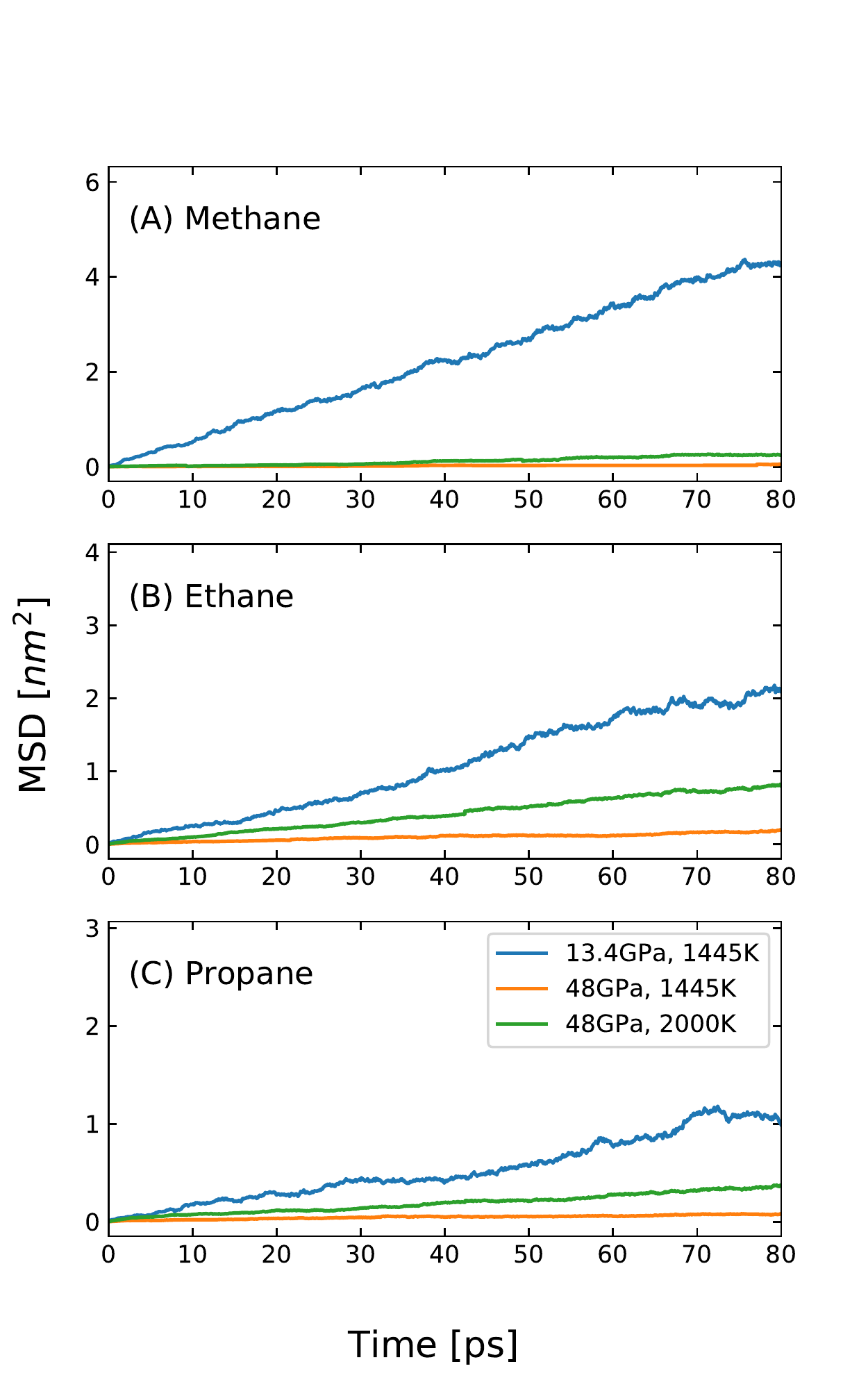}
    \caption{Mean squared displacements of carbon atoms in (A) methane, (B) ethane, and (C)propane at extreme P-T conditions.}
    \label{MSD}
    \end{figure}

Fig. \ref{MSD} shows the mean squared displacements of methane, ethane, and propane at 13.4 GPa, 1445 K; 48 GPa, 1445 K; and 48 GPa, 2000 K. 
All the alkanes are in the supercritical state, except methane at 48 GPa and 1445 K, which is an amorphous solid.
It is interesting to see that methane freezes at 1445 K, whereas propane does not. 
At ambient pressure, the melting point of methane is 90.7 K, while propane melts at 85.5 K \cite{thalladi2000melting}. In fact, the melting point of propane is unusually lower than other n-alkanes,
which may be due to the low packing efficiency of propane molecules \cite{thalladi2000melting}.

In our simulations, a few C-H bonds broke temporarily, 
but we did not see any new chemical bond form.
Experimentally, Nellis et al. reported no chemical reactions in shock wave experiments 
up to 26 GPa and 3200 K \cite{nellis_electrical_2001}, whereas Lobanov et al. found that heavier alkanes and unsaturated hydrocarbons may form  in laser-heated diamond anvil cells above 1500 K \cite{lobanov_carbon_2013}.
Theoretically, 
Spanu et al. found that the mixtures of higher hydrocarbons and hydrogen become thermodynamically more stable than methane at above 4 GPa and 1000$\sim$2000 K using AIMD simulations and free energy calculations, but without metal catalysts or unsaturated diamond surfaces they did not see chemical reactions in that P-T range \cite{spanu_stability_2011}.
Thus, we conclude that activation barriers of alkanes may be too high,
so we could not see chemical reactivity within limited simulation time.

\begin{figure}
    \centering
    \includegraphics[width=0.6\textwidth]{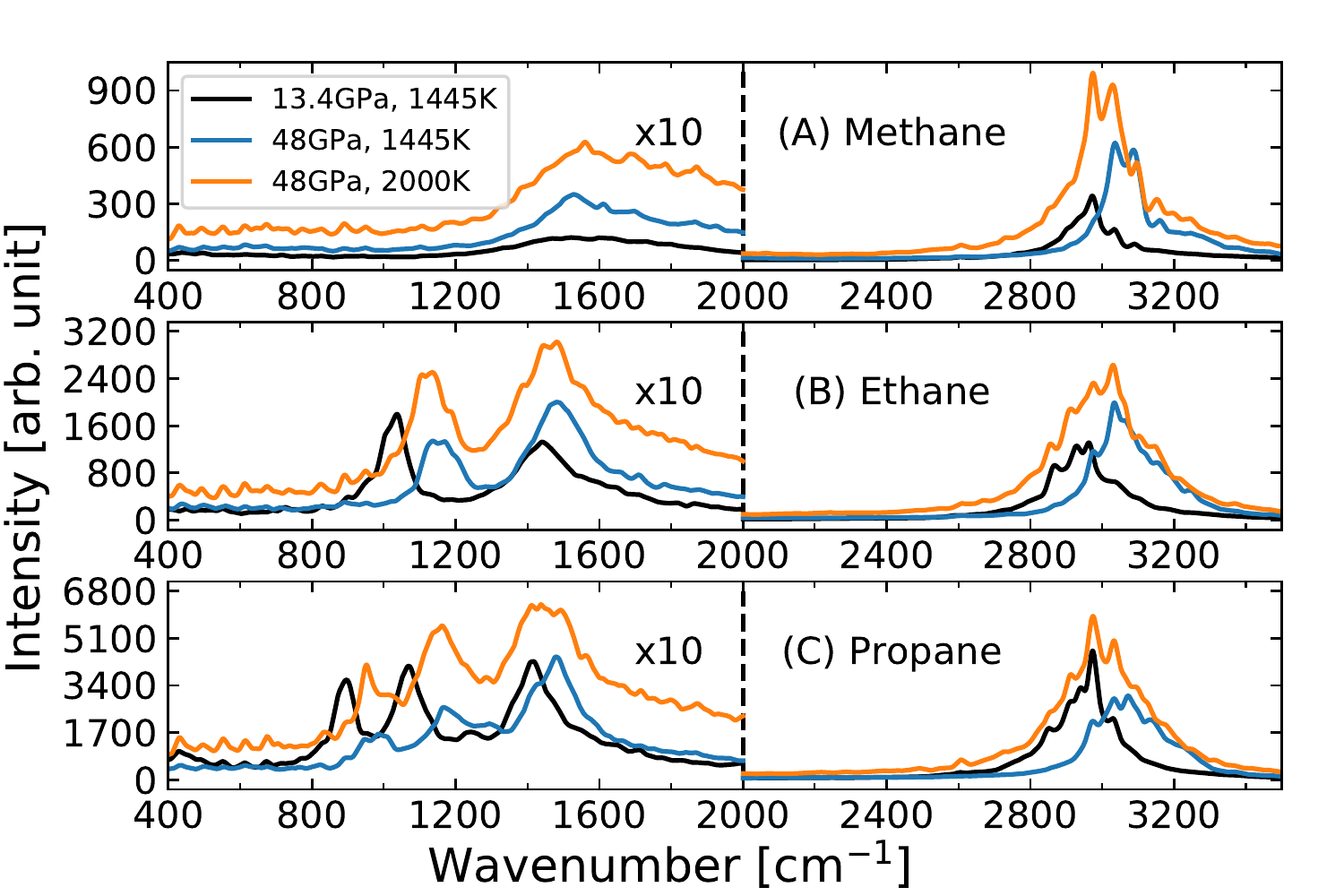}
    \caption{Unpolarized Raman spectra of (A) methane, (B) ethane, and (C) propane at extreme P-T conditions.
    The Raman intensity is multiplied by 10 below 2000 cm$^{-1}$.}
    \label{whole-raman}
\end{figure}

Fig. \ref{whole-raman} shows the unpolarized Raman spectra of hydrocarbons, from which we know how Raman bands change with pressure and temperature. 
The C-H bond stretching region is between 2800 and 3200 cm$^{-1}$.
The stretching band downshifts with increasing temperature or decreasing pressure.
For methane at 48 GPa and 1445$\sim$2000 K,
the stretching band splits into two peaks,
which was found as an indicator of orientational ordering of molecules in compressed methane crystals at least up to 30 GPa \cite{bini_high_1995,bini_high-pressure_1997}, suggesting that CH$_4$ molecules may have some orientational correlation under such extreme P-T conditions.

\begin{figure}
  \centering
  \vspace{5mm}
  \includegraphics[width=0.5 \textwidth]{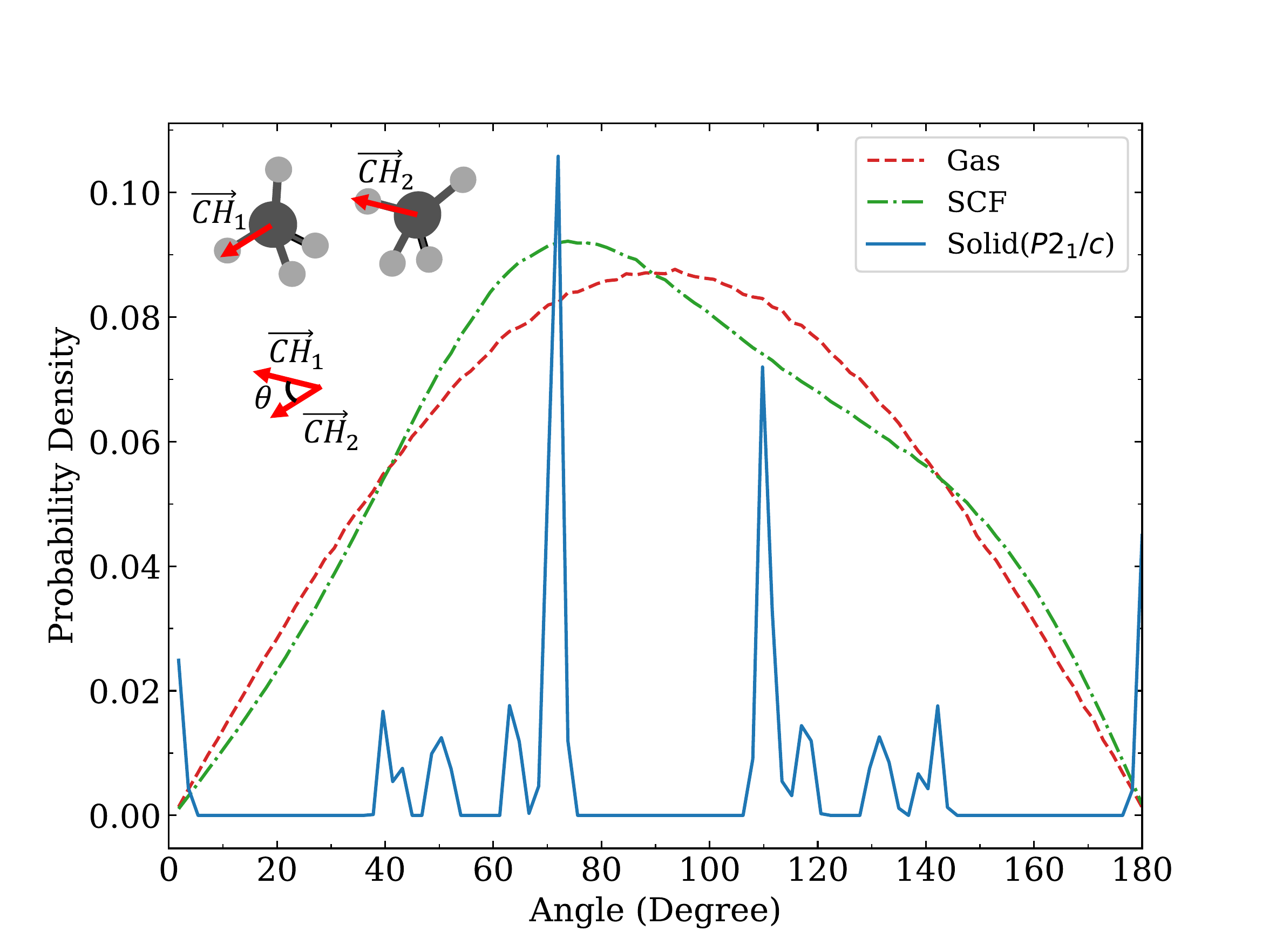}
  \caption{Angle distribution of two C-H bonds respectively from two nearest methane molecules.
  Three methane phases are compared: gas, supercritical fluid (SCF) at 48 GPa and 2000 K, and solid (phase group: $P2_1/c$).
  The probability densities of the gas and liquid phases are multiplied by 10. }
  \label{angle-dist}
  \end{figure}

Fig. \ref{angle-dist} shows the angle distribution of two C-H bonds respectively from two nearest methane molecules.
For the crystalline methane (space group: $P2_1/c$ \cite{lin_structural_2011}), the angle distribution has two main peaks at 70.5$^{\circ}$ and 109.5$^{\circ}$, and the former peak is taller than the latter one.
In gas phase, the angle is randomly distributed, so the probability density function is $\frac{\pi}{360^{\circ}}\sin\theta$. 
At 48 GPa and 2000 K, the angle distribution becomes asymmetric, different from the $\sin\theta$ function, indicating a certain angular correlation between neighboring molecules.

\begin{figure}
    \centering
    \includegraphics[width=1.0\textwidth]{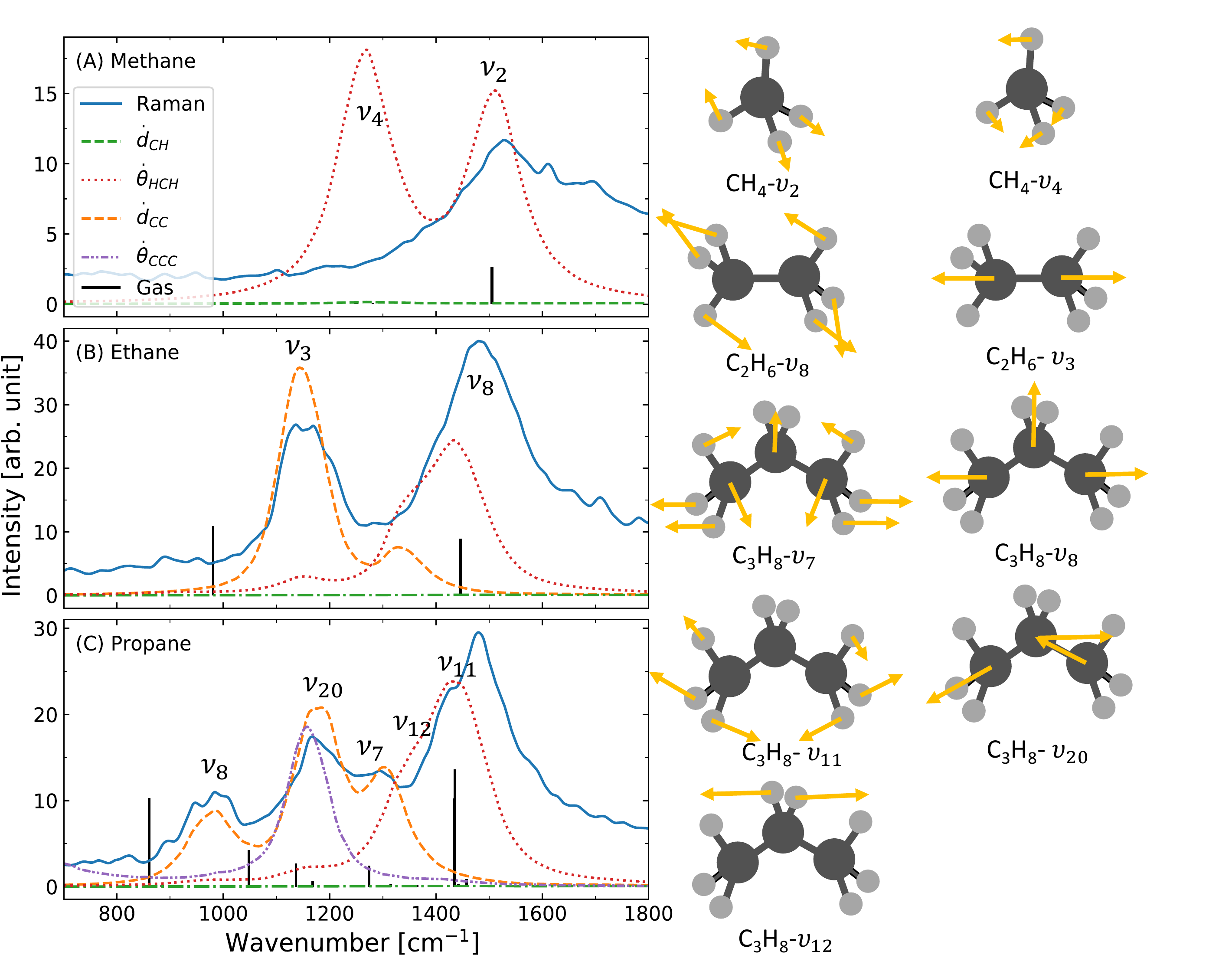}
    \caption{ Raman band assignments.
    The Raman, C-H bond ($d_{CH}$), H-C-H angle ($\theta_{HCH}$), C-C bond ($d_{CC}$), C-C-C angle ($\theta_{CCC}$) spectra are compared.
    The Raman spectra of methane, ethane, and propane are multiplied by 0.333, 0.200, and 0.067, respectively. The pressure is 48 GPa and the temperature is 1445 K.}
    \label{peak-recognition}
\end{figure}

The C-H stretching bands of hydrocarbons largely overlap, whereas
the Raman bands in the low frequency region between 600 and 2000 cm$^{-1}$ show distinct peaks for methane, ethane, and propane, so the Raman modes in this region are often used to identify hydrocarbons in experiment \cite{buldakov_raman_2013}. 
Methane only shows one peak between 1400 and 1600 cm$^{-1}$, which upshifts with increasing temperature.
For an isolated methane molecule, 
the calculated Raman active mode in this range
appears at $\sim$1505 cm$^{-1}$, which corresponds to the 
degenerate deformation $\nu_2$ \cite{shimanouchi1973tables}. 
We also performed the Fourier transform of autocorrelation functions of C-H bonds and H-C-H angles, respectively:
\begin{align}
A_{CH}(w)=\int d t e^{-i w t}\frac{\left\langle\dot{d}_{C H}(0) \dot{d}_{C H}(t)\right\rangle}{\left\langle\dot{d}_{C H}(0) \dot{d}_{C H}(0)\right\rangle} \\
A_{HCH}(w)=\int d t e^{-i w t}\frac{\left\langle\dot{\theta}_{HC H}(0) \dot{\theta}_{HC H}(t)\right\rangle}{\left\langle\dot{\theta}_{HC H}(0) \dot{\theta}_{HC H}(0)\right\rangle} ,
\end{align}
where $d_{CH}$ is the C-H bond length
and $\theta_{HCH}$ is the H-C-H bond angle.
As shown in Fig. \ref{peak-recognition}(A), 
there is no C-H stretching mode between 1000 and 1800 cm$^{-1}$.
The H-C-H angle spectrum has two peaks:
$\nu_4$ at 1269 cm$^{-1}$ and $\nu_2$ at 1511 cm$^{-1}$. 
We can not identify the $\nu_4$ mode in the calculated Raman spectrum in Fig. \ref{peak-recognition}(A), 
because the Raman cross section of $\nu_2$ is about 28 times as much as that of $\nu_4$ for an isolated CH$_4$ molecule.

Fig. \ref{peak-recognition} (B) shows the Raman spectra of ethane between 800 and 1800 cm$^{-1}$. 
For an isolated ethane molecule, there are two notable Raman modes in our calculation: $\nu_3$ at 987 cm$^{-1}$ and $\nu_8$ at 1455 cm$^{-1}$, corresponding to the C-C stretching and CH$_3$ degenerate deformation modes, respectively \cite{shimanouchi1973tables}.
At 48 GPa and 1445 K, both of the peaks upshift.
In addition to the C-H bond and H-C-H angle spectra, we also calculated the C-C bond spectrum:
\begin{equation}
A_{CC}(w)=\int d t e^{-i w t}\frac{\left\langle\dot{d}_{C C}(0) \dot{d}_{C C}(t)\right\rangle}{\left\langle\dot{d}_{C C}(0) \dot{d}_{C C}(0)\right\rangle} ,
\end{equation}
where $d_{CC}$ is the C-C bond length.
It has a main peak at 1145 cm$^{-1}$, which overlaps with the $\nu_3$ band, indicating that the $\nu_3$ mode is indeed the C-C stretching mode.
The $\nu_8$ mode is found in the vicinity of the H-C-H angle spectrum, suggesting that the CH$_3$ degenerate deforming is relevant to the variation of H-C-H angles.
Comparing Fig. \ref{peak-recognition} (A) with (B), we found that
the $\nu_2$ band of methane overlaps with the $\nu_8$ band of ethane, 
and the $\nu_3$ band of ethane is missing in the Raman spectra of methane, so this band can be used as a characteristic signal for the C-C bond formation under high P-T conditions. 
The $\nu_3$ mode upshifts with increasing pressure, and downshifts with increasing temperature.
The pressure effect is more obvious than the temperature effect in the P-T range studied here, as shown in Fig. \ref{whole-raman}.

Fig. \ref{peak-recognition} (C) shows the Raman spectra of propane between 800 and 1800 cm$^{-1}$. 
In our calculation, an isolated propane molecule has notable Raman modes as follows:
C-C stretching mode $\nu_8$ at 860 cm$^{-1}$,
C-C stretching mode $\nu_{20}$ at 1048 cm$^{-1}$,
CH$_{3}$ rocking mode $\nu_7$ at 1137 cm$^{-1}$,
CH$_{2}$ twisting mode $\nu_{12}$ at 1274 cm$^{-1}$, and
CH$_{3}$ degenerate deforming mode $\nu_{11}$ at 1435 cm$^{-1}$ \cite{shimanouchi1973tables}.
At 48 GPa and 1445 K, we can identify the $\nu_{12}$ and $\nu_{11}$ modes in the H-C-H angle spectrum.
The C-C bond spectrum has three peaks: $\nu_8$, $\nu_{20}$, and $\nu_7$. 
We also calculated the C-C-C angle ($\theta_{CCC}$) spectrum:
\begin{equation}
    A_{CCC}(w)=\int d t e^{-i w t}\frac{\left\langle\dot{\theta}_{CCC}(0) \dot{\theta}_{CCC}(t)\right\rangle}{\left\langle\dot{\theta}_{CCC}(0) \dot{\theta}_{CCC}(0)\right\rangle} ,
\end{equation}
which has one peak overlapping with the $\nu_{20}$ mode, indicating that this CC stretching mode also considerably changes the C-C-C bond angle.
The $\nu_8$ band does not show up in the spectra of methane and ethane, so it can be used to verify the formation of the C-C-C bonds.

We further divided Raman spectra into inter- and intramolecular contributions with the help of maximally localized Wannier functions \cite{gygi_computation_2003}, which provide local molecular orbitals to partition electron density into molecules \cite{marzari_maximally_2012}. 
The effective molecular polarizability, $\alpha_{mol}^{eff}$, was obtained by:
\begin{equation}
    \alpha_{mol}^{eff} \vec{E} = -e\int_\Omega \vec{r}\Delta \rho_{mol} d\vec{r},
\end{equation}
where $\vec{E}$ is the macroscopic electric field,
$e$ is the elementary charge, $\Delta \rho_{mol}$
is the electron polarization density of the molecule induced by the electric field, and the integral is over the whole simulation box ($\Omega$).

The autocorrelation functions in Eqs. (\ref{eq:isotropic}) and (\ref{eq:aninstropic}) are respectively written as,
\begin{align}
    %\begin{aligned}
    \langle\bar{\alpha}(0) \bar{\alpha}(t)\rangle
    &=\left\langle\sum_{\genfrac{}{}{0pt}{2}{m=1}{m \neq n}}^{N_{\mathrm{mol}}} \sum_{n=1}^{N_{\mathrm{mol}}} \bar{\alpha}_{\mathrm{mol}, m}^{\mathrm{eff}}(0) \bar{\alpha}_{\mathrm{mol}, n}^{\mathrm{eff}}(t)+\sum_{m=1}^{N_{\mathrm{mol}}} \bar{\alpha}_{\mathrm{mol}, m}^{\mathrm{eff}}(0) \bar{\alpha}_{\mathrm{mol}, m}^{\mathrm{eff}}(t)\right\rangle
    %\end{aligned}
    \label{Eq:inter-intra-alpha}\\
%
    %\begin{aligned}
    \langle \Tr(\beta(0) \beta(t))\rangle
    &=\left\langle  \sum_{\genfrac{}{}{0pt}{2}{m=1}{m \neq n}}^{N_{\mathrm{mol}}} \sum_{n=1}^{N_{\mathrm{mol}}} \Tr(\beta_{\mathrm{mol}, m}^{\mathrm{eff}}(0) \beta_{\mathrm{mol}, n}^{\mathrm{eff}}(t))+\sum_{m=1}^{N_{\mathrm{mol}}} \Tr(\beta_{\mathrm{mol}, m}^{\mathrm{eff}}(0) \beta_{\mathrm{mol}, m}^{\mathrm{eff}}(t)) \right\rangle
    %\end{aligned}
    \label{Eq:inter-intra-beta}
\end{align}
where $N_\mathrm{mol}$
is the total number of hydrocarbon molecules.
We applied the Fourier transform on
the first and second terms of Eqs.
(\ref{Eq:inter-intra-alpha})(\ref{Eq:inter-intra-beta})
to get the inter- and intramolecular 
Raman spectra, respectively.

\begin{figure}
    \centering
    \includegraphics[width=0.6\textwidth]{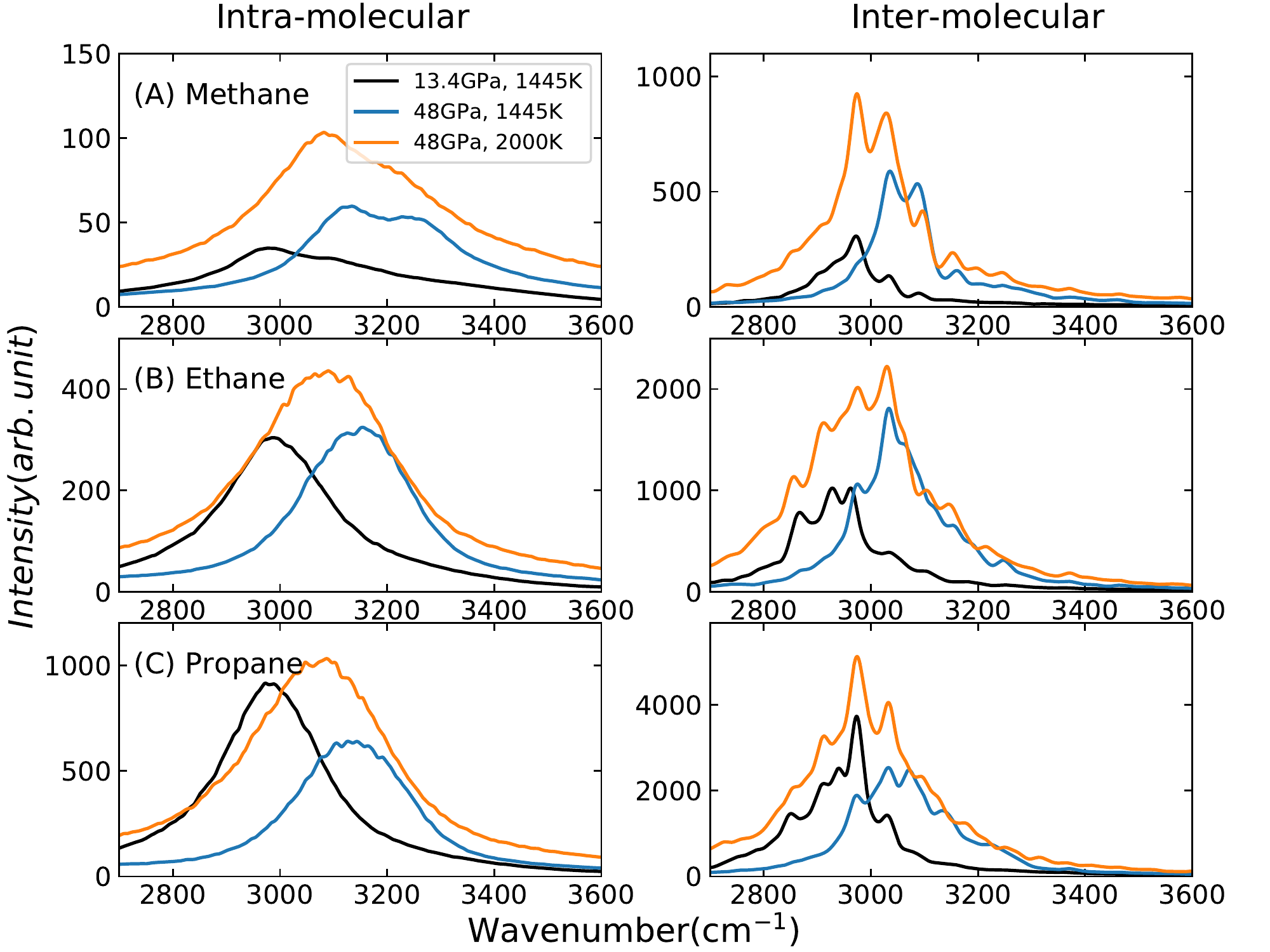}
    \caption{Intramolecular ( left panel ) and intermolecular ( right panel ) contributions
    to unpolarized Raman spectra at extreme P-T conditions: (A) methane, (B) ethane, and (C) propane.}
    \label{inter-intra-raman-high}
\end{figure}

Fig.~\ref{inter-intra-raman-high} shows the inter- and intramolecular Raman spectra in the high frequency region.
The intensities of intermolecular contributions are much larger than the intramolecular ones, indicating strong molecular interactions under extreme P-T conditions.
The intramolecular spectra have only one main peak, whereas
there appears to be a few peaks in the intermolecular spectra,
which may come from multiple intermolecular couplings between hydrocarbon molecules.

\section{Conclusion}
In summary, we calculated the unpolarized Raman spectra of methane, ethane and propane at 13.4$\sim$48 GPa and 1445$\sim$2000 K using ab initio molecular dynamics simulations coupled with density functional perturbation theory. Our method considers the Raman selection rule and anharmonic temperature effects.
We monitored how pressure and temperature change the Raman bands, and understood the Raman bands at the molecular scale. 
Particularly, we identified the characteristic Raman bands for the formation of C-C and C-C-C bonds. 
We found that the C-H stretching bands of supercritical hydrocarbons in the high-frequency region split due to the intermolecular interactions.
Our results may help to interpret in-situ Raman measurements of hydrocarbons at extreme P-T conditions, which are of great importance in understanding the deep carbon cycle inside Earth and the compositions of ice giant planets.

\section{acknowledgments}
\begin{acknowledgments}
D.P. acknowledges support from the Croucher Foundation through the Croucher Innovation Award, Hong Kong Research Grants Council (Projects ECS-26305017 and GRF-16307618), National Natural Science Foundation of China (Project 11774072 and Excellent Young Scientists Fund), and the Alfred P. Sloan Foundation through the Deep Carbon Observatory.
Part of this work was carried out using computational resources from the National Supercomputer Center in Guangzhou, China.
\end{acknowledgments}

\bibliography{main}

\newpage

\end{document}